# High Throughput Oxide Lattice Engineering by Parallel Laser Molecular Beam Epitaxy and Concurrent X-ray Diffraction

M. Ohtani

Department of Physics and Institute for Material Research, Tohoku University, Sendai 980-8577, Japan

M. Lippmaa, and T. Ohnishi

Institute for Solid State Physics, University of Tokyo, Kashiwa, 277-8581, Japan

M. Kawasaki[a]

Institute for Materials Research, Tohoku University, Sendai 980-8577, Japan and Combinatorial Materials Exploration and Technology (COMET), National Institute for Material Science (NIMS), Tsukuba, 305-0044, Japan

Abstract

A novel laser molecular beam epitaxy (LMBE) system for the fabrication of atomically controlled oxides superlattices and an x-ray diffractometer that measures spatially-resolved x-ray diffraction spectra have been developed based on the concept of combinatorial methodology. The LMBE chamber has two moving masks, an automated target stage, a substrate heating laser, and an in-situ scanning reflection high-energy electron diffraction system. The x-ray diffractometer with a curved monochromator and two-dimensional detector is used for rapid concurrent x-ray diffraction intensity mapping with the two axes of the detector corresponding to the diffraction angle and a position in the sample.

---

[a] Electronic mail: kawasaki@imr.tohoku.ac.jp

I. INTRODUCTION

Artificial oxide superlattices have recently attracted much attention, since properties of oxides can be tailored using superlattice effects such as quantum confinement, spin exchange, charge transfer, and epitaxial strain. There are many reports on oxide superlattices concerning their optical[1] and transport[2] properties, which are also known in semiconductor superlattices. Additionally, recent advances in epitaxial growth techniques of oxide thin films and effective surface termination[3] control have extended oxide superlattice studies to magnetic[4] and ferroelectric materials[5]. Not only the growth of conventional 'bicolor' superlattices composed of two components, but also the growth of 'tricolor' superlattices, which are composed of three kinds of oxides and stacked as ABCABC, have been reported recently[6,7,8,9]. However, there is no general theoretical guideline that could be used to predict the properties of an oxide superlattice. This is largely due the wide choice of elements, covering most of the periodic table, that can be included in structurally similar oxides and thus integrated into a superlattice. Besides, the optimization of growth parameters for oxide superlattices is a very time-consuming process because the oxide crystal growth is very sensitive to process conditions, e.g. substrate temperature, oxygen pressure, etc. All of these parameters may need to be optimized for each individual material used in a superlattice. For these reasons, conventional one-by-one synthesis is very inefficient. On the other hand, combinatorial methodology, which is well established in the area of drug development and organic synthesis, has been adapted to the growth of libraries of thin films.[10] Our purpose was to use this methodology for high-throughput synthesis of oxide superlattices.

The original combinatorial solid–state synthesis process used by Xiang et al.,

consists of three basic steps.[10] First, a large number of amorphous thin films with different compositions are deposited at room temperature on a single substrate in a single deposition run in the form of discrete cells or continuously varying composition gradients. Second, the library is annealed to turn the deposited material into homogenous amorphous films. Finally, the library is sintered at high temperature in order to crystallize the films. Although this method is very efficient for obtaining thermodynamically stable materials, the use of a sintering step that is required to crystallize amorphous precursors, precludes the use of this technique for the growth of combinatorial superlattice libraries. To overcome this difficulty, a non-equilibrium epitaxial growth process has to be employed. Thus various groups have developed laser molecular beam epitaxy (LMBE) systems to fabricate combinatorial libraries of epitaxial thin films[11,12,13,14]. However these systems cannot observe reflection high energy electron diffraction (RHEED) specular spot intensity variations at several points on the substrate simultaneously. We believe that combinatorial synthesis of atomically controlled oxide thin films using scanning RHEED becomes more important as we start applying high-throughput synthesis techniques to more complex crystal structures.

Here, we have developed a LMBE system to synthesize atomically controlled combinatorial libraries[15,16] and a high throughput x-ray diffraction technique for characterizing the crystal structure of epitaxial thin film libraries[17,18]. A combination of these two techniques can be used to achieve significant increases in the throughput of developing and optimizing oxide superlattices.

II. COMBINATORIAL LASER MBE

A. Outline

In order to materialize the concept of combinatorial epitaxial crystal lattice integration, we have constructed a novel LMBE system, which is illustrated in Fig. 1. The features of the system are: an automated 4-target stage, masking system with moving stencil and shadow masks, in-situ scanning RHEED system[15,19], and a substrate heating system using an infrared laser[20]. Target exchange, mask movement, programming of deposition sequences, and reaction diagnostics are all controlled by a computer. The control software is implemented in LabVIEW$^{TM}$, which provides a user-friendly graphical programming environment that is well-proven for control and measurement applications[21]. The combinatorial laser MBE system enables us to fabricate various types of combinatorial samples, as shown in Fig. 2. By using two-dimensional binary masks, it is possible to fabricate combinatorial libraries with discrete cells, e.g. a 10 x 10 array (Fig. 2(a)). One-dimensional superlattice arrays can be fabricated by using a one-dimensional mask and scanning RHEED (Fig. 2(b)) [15,22]. Temperature-gradient films are fabricated by using a specially designed sample holder and a tightly focused heating laser (Fig.2 (c))[23,24]. By synchronizing the motion of the moving masks during deposition with the triggering of the deposition laser, one can fabricate atomically mixed, epitaxial composition-spread films (Fig. 2(d))[25]. This technique can also be extended to ternary composition-spread growth by allowing for automated sample rotation.[26,27] It is possible to combine various methods, e.g. to obtain two-dimensional parallel film libraries in which the growth temperature varies along one direction and the film composition varies along another direction on the substrate surface.[28,29]

B. System description

Figures 3(a) and (b) show the outside and inside views of a combinatorial LMBE

chamber. The growth chamber is built around a 260 mm diameter sphere to minimize the size of the system. The top CF203 flange holds the sample stage, heating laser optics, and two mask manipulators. The distance between the stencil mask plate and the sample surface is about 1 mm. The sample stage can be rotated for sample exchange, ternary deposition modes, and for acquisition of RHEED images. The bottom flange of the chamber has an automated ablation target stage. Four ablation targets are mounted on the stage with double rotary feedthroughs for continuous target rotation and target selection. In order to give a clear view of the sample surface for a mask alignment camera or a pyrometer, the target stage can be moved away from the center of the chamber, giving a direct line of sight at the sample from a viewport at the center of the bottom flange. The same viewport can be used for in situ optical characterization of thin film libraries. A pulsed Compex 102 KrF excimer laser (Lambda-Physik) was used for ablation. The laser light was introduced into the chamber at a 45 degree angle to the target surface through a fused-silica window. A compact semiconductor diode laser, Jenoptik JOLD-120-CAXF-5A, with a maximum continuous wave power of 120 W at a wavelength of 808 nm was used for heating the sample. Although we have used Nd-YAG lasers for sample heating in earlier systems, we have found the semiconductor lasers to be more stable and cheaper than the industrial-grade Nd-YAG lasers. Even at 120 W, the optical power is sufficient to heat a 10 mm × 10 mm substrate up to 1400 ºC. Light from the heating laser is delivered via an optical fiber with a 600 μm core diameter and focused on the backside of an oxidized Nickel or stainless steel sample holder by simple two-lens optics. The pyrometer detection wavelength was 2 μm and had a measurement range of 180 to 1400 °C. The measurement spot size on the sample surface was about 2 mm in diameter. The oxygen gas pressure was controlled with a manual variable leak valve. To obtain a high partial pressure for the oxidization of the film surface, a nozzle tube for oxygen gas was set at about 100 mm distance form the substrate. In order to monitor the sample and the masks, we used two cameras, one located at a viewport at the center of the bottom flange, having a direct view of the sample when the target stage was moved to the side, and another at an angle of 30º to the normal at a side port of the chamber for monitoring the sample and mask movement during deposition.

C. Electronics and software

A schematic of the system is shown in Fig. 4. Sample rotation, movement of two masks, target carrousel rotation, and the target stage flip motion were controlled via two 4-axis motor controllers from National Instruments, model PCI-7314. The power of the excimer laser was controlled via an RS-232C serial link. The vacuum gauge controller was also attached to the computer via a serial link. Automatic pressure control is possible when a motorized variable leak valve is used. Since this is rarely necessary, most of our chambers, including the system described here, use manual leak valves. In order to perform PID temperature control of the sample, the heating laser power was controlled and the substrate temperature monitored using an optical pyrometer via a multifunction data acquisition board, National instruments PCI-6035E. This system compensated automatically for slow laser power drift and for temperature-dependent or ambient pressure dependent sample heating efficiency changes. The sample temperature could be changed at rates of up to about 100 °C/min using PID control. Constant laser output power was used when significant changes in the sample emissivity were expected, for example when starting a heteroepitaxial deposition. The system also monitors turbo pump power, heating laser power, and laser cooling via the multifunction data acquisition board for maintenance purposes. A real-time H8-series microprocessor from Renasas was used to generate control signals for the video cameras and to synchronize the scanning RHEED camera with deposition laser trigger pulses. All chamber function can be controlled from a control panel written in LabVIEW$^{TM}$. This control program has a scripting interface that is used to perform automated control tasks. Using the script, we can automate the film growth procedure, such a composition spread and parallel superlattice synthesis. The built-in multitasking capabilities of LabVIEW$^{TM}$ were used to obtain a control program that can run several autonomous parts of the system concurrently. A guiding principle of the software design was to always allow interactive override or control of all system parameters, even when executing automated sequences. This makes it easier for the operator to make small alterations or modifications during the deposition.

D. Scanning RHEED

In order to control the thicknesses of individual superlattice layers with atomic layer accuracy, film growth has to be monitored at several points on the substrate simultaneously during combinatorial superlattice synthesis. The combinatorial LMBE system was equipped with a differentially-pumped RHEED gun with two sets of beam deflection coils, a camera for RHEED image acquisition and a deflection coils driver controlled by computer. The scanning RHEED system was driven by an embedded real-time processor that ensured synchronization between electron beam movement, image acquisition, and deposition laser triggering. The RHEED pattern and intensity monitoring software is also written in LabVIEW$^{TM}$ and can run concurrently on the same PC that is used for controlling the chamber functions. In order to measure film growth concurrently at many points on the sample surface, the electron beam can be moved to a new position for each video frame of the image acquisition camera. The beam movement occurs during the blanking period between video frames and image acquisition and feature analysis is performed at the full video rate.

The electron beam deflection system used two coils for vertical beam motion and another two coils for horizontal beam motion. In the case of a conventional RHEED system with one coil set, the electron beam tilt generated by the deflection coils cannot be compensated and the incident angle of the electron beam on the sample surface depends on which point is being measured. For a typical 10 mm sample this incident angle deviation can reach 1 degree when moving the electron beam from the center of the sample to the edge of the substrate. For RHEED image observation, however, we need to maintain the incident electron beam parallel to a selected high-symmetry direction with an accuracy of better than 0.1 degrees, in order to avoid changes in the diffraction pattern. Using two deflection coil sets, it is possible to avoid this problem and obtain true parallel electron beam sweeping for each selected position in a combinatorial library. For demonstrating the performance of the system, we show the results of an experiment where three $SrTiO_3$/$BaTiO_3$ superlattices with different periodicities were fabricated on a single $SrTiO_3$ (001) substrate at an oxygen pressure 1 x $10^{-6}$ Torr and substrate temperature of 700 ºC. Figure 5 shows the observed RHEED intensity variation monitored at three regions. The deposition time for each region was determined by counting the RHEED oscillations in each region. The deposition sequence and mask position are shown at the bottom of Fig. 5. Using this process, we can fabricate ten atomically controlled superlattices on a single substrate in parallel.

The same deflection system can also be used to observe simultaneously RHEED specular spot intensity variations at different vertical incident angles. Choosing the incident angle is very important to observe RHEED oscillations because the contrast between maximum and minimum of RHEED intensity depends on the incident angle. Traditionally, the sample height or tilt is changed in order to change the incident angle, but the angle has to be set before the experiment starts. Changing the sample height also changes the sample to target distance and thus the characteristics of the ablation plume that hits the sample surface during deposition. We can eliminate this problem with the scanning beam system and observe RHEED oscillations at various angles simultaneously. We can then select the most suitable data set after the experiment. An additional advantage is that the same data can be used to construct crude RHEED rocking curves as a function of time or surface coverage. This can give useful information about the dynamics of surface changes during the initial stages of heteroepitaxial film growth. The electron beam geometry in such experiments is shown in Fig. 6(a). When the electron beam is bent upward by the first coil set and downward by the second coil, we obtain a very low incident angle. When the electron beam is bent downward by the first coil set and upward by the second coil, we obtain a large incident angle. A typical angle range covered by our chambers is 0 to 4 degrees. Figure 6(b) shows RHEED intensity oscillations during homoepitaxial growth on a $SrTiO_3$ (001) substrate at five different angles: 0.6º, 0.8 °, 1.1 °, 1.4 °, and 1.7 °. Among these angles, 0.6º, 1.1 °, and 1.7 ° meet Bragg conditions (on-Bragg), and 0.8º and 1.4º are off-Bragg. RHEED intensity oscillations at low incident angles (0.6º and 0.8 °) show phase inversion. The data obtained at 1.7 ° is probably the most useful set for measuring the film thickness.

III. CONCURRENT X-RAY DIFFRACTOMETER

For the high-throughput characterization of combinatorial epitaxial thin films for laboratory use, we have developed an x-ray diffractometer (XRD), which we call concurrent XRD (CXRD).

A. Outline and principle

Figure 7(a) shows a schematic diagram of the concurrent XRD beam geometry. An x-ray beam is generated from a fine-focus x-ray generator. In order to eliminate Cu K$\alpha_2$, Cu K$_\beta$, and white radiation and to focus the x-rays into an area of ~10 mm × 100 μm on the sample surface with a convergence angle of 2°, a curved monochromator was used. A 4-axes sample stage and a CCD camera were mounted on the $\omega$ and $2\theta$ stages of a conventional two-circle goniometer, respectively. The diffracted x-rays from the sample are acquired with a CCD camera. The horizontal axis in the image corresponds to the $2\theta$ diffraction angle and the vertical axis in the image is a simple one-to-one mapping to the position on the sample surface. The intensity in each horizontal line of the image thus corresponds to a $2\theta$ diffraction pattern, measured at a particular point along the line illuminated by the x-rays on the sample surface. Figure 7(b) shows an Ewald construction in reciprocal space for the concurrent XRD configuration. The shaded square ($\square$ABCD), defined by the four Ewald spheres is examined in the measurement. The size of the square is determined by the convergence angle of the x-rays exiting the curved monochromator. The x-ray intensity variation along the $2\theta$ axis on the CCD camera corresponds to an $\omega$-$2\theta$ XRD pattern. In this system, however, the CCD camera cannot distinguish between different incident angles of x-rays impinging on the sample. For instance, x-rays coming from different $\omega$ angles as represent by 1, 2, and 3 in Fig. 7(b) can contribute to the intensity detected at the point 1'. The x-ray path from 1 through the sample to 1' corresponds to the reciprocal space point A. The path from 3 to 1' corresponds to B. Therefore, the intensity recorded at a $2\theta$ angle of 1' is an integral of x-ray intensities along the arrow extending from reciprocal space point B to point A. Projections of reciprocal lattice points in the square along the arrows BA, PQ, and CD are recorded as the diffraction pattern by the CCD camera. The spatial resolution (~ 100 μm) and diffraction angle resolution (~ 0.025º) are limited by the diameter of incident x-rays and the pixel size of the CCD camera, respectively. The measurement time varies from five seconds to a few minutes.

B. Description of the system

An image of the concurrent XRD system is shown in Fig. 7 (c). As an x-ray generator, we use a Rigaku Ultra-18 rotating anode Cu tube with a maximum power of 1.2kW. The effective focus size is 0.1 mm x 0.1 mm. The x-ray beam passed through a Johan-type monochromator, utilizing α-quartz ($10\bar{1}1$) plane diffraction at an offset angle of 3º and 2R = 747 mm. The monochromator was adjusted to homogenize the x-ray intensity in lateral direction by the four-point bending method. The x, y, z motion of the sample stage, ω, θ, φ axes and the z motion of the camera stage are driven by a stepping motor controller, PM16C-02N-OP from Tsuji Denshi. We used a Hamamatsu C4880-20-24WF CCD with a fluorescent screen as a two-dimensional x-ray detector. This CCD camera is cooled to -30ºC by a Peltier element and water circulators. The x-ray source to monochromator, monochromator to focusing point, and focusing point to camera distances are 210 mm, 134 mm, and 350 mm, respectively. The diffraction image is acquired via a National Instruments PCI-1424 digital image acquisition board.

The x-ray shutter is synchronized with image acquisition. The motor control and image acquisition is controlled by LabVIEW$^{TM}$. Using a script interface, we can easily do repeated measurements like XRD mapping[17]. Analytical software was developed using the IGOR$^{TM}$ data analysis package.

C. Examples

1. Composition-spread epitaxial thin films

As a demonstration of the capabilities of the CXRD system, we studied $(Ba_x,Sr_{1-x})TiO_3$ ($0 \leq x \leq 1$) epitaxial composition-spread films grown on $SrTiO_3$ (001) substrates by combinatorial LMBE. Ceramic targets of $BaTiO_3$ and $SrTiO_3$ were ablated alternately by switching the targets, while the moving masks were slid synchronously to give a linear gradient of $x$ in $(Ba_x,Sr_{1-x})TiO_3$. The deposition temperature and oxygen pressure during the growth were 700ºC and $10^{-5}$ Torr, respectively. The schematic of the composition-spread film is shown at the top of Fig. 8(a). Figure 8(b) shows the diffraction image obtained from the film. Diffraction features relating to the (002) plane of the substrate and the composition-spread film are annotated in the Fig. 8(b). The vertical and the horizontal axes correspond to the $2\theta$ diffraction angle and the film composition $x$ (*i.e.* position in the film), respectively. Color denotes the intensity of diffracted x-rays. The horizontal high-intensity line at 46.5° corresponds to the (002) substrate peak. The adjacent curve-shape diffraction feature corresponds to the composition-spread film. Figure 8(c) shows the out-of-plane and in-plane lattice constants of the composition-spread film. The out-of-plane lattice constant can be calculated from the diffraction angle of the peak in a straightforward fashion and the in-plane lattice constant can be evaluated by using two additional asymmetric diffraction images taken in e.g. (013) and ($0\bar{1}3$) directions. The details of this procedure can be found in Ref. 18. For $x \leq 0.6$, the diffraction peak of the film is sharp and the out-of-plane lattice constant increases while the in-plane lattice constant stays almost identical to that of the substrate with increasing $x$. For $x > 0.6$, the peak becomes broad and the out-of-plane lattice constant abruptly decreases at $x \sim 0.6$ and remains almost constant for larger $x$, whereas the in-plane lattice constant abruptly increases at $x \sim 0.6$ and gradually increases with increasing $x$. The results mean that the compressive stress of the film is abruptly relaxed at $x = 0.6$ as can be clearly seen by the abrupt change of the in-plane and out-of-plane lattice constants. It is noted that the evolution of relaxation in the strained lattice can be visualized as a function of the degree of lattice mismatch. We can determinate a critical point for the lattice relaxation by only one composition-spread film and three concurrent XRD images. The whole process takes less than one day.

2. Combinatorial superlattice library

Figure 9(a) shows a concurrent XRD image of ten superlattices $[(SrTiO_3)_n/(BaTiO_3)_n]_{30}$ ($n$ = 12, 14, ~, 30), integrated on a single $SrTiO_3$ (001) substrate as illustrated in Fig. 9(b). The sample was synthesized by monitoring in-situ the intensity oscillation of scanning RHEED for all the superlattice pixels simultaneously with a combinatorial LMBE system. The line-shaped diffraction peak at around $2\theta = 23°$ in Fig. 9(a) is the (001) diffraction peak from the $SrTiO_3$ substrate. The adjacent lines with steps correspond to the fundamental and satellite peaks of each superlattice. The $2\theta$ values of the fundamental peaks do not shift since the average lattice constants of the superlattices do not change. The systematic shift of satellite peaks corresponds to the change in superlattice periodicities. By taking cross sections of the data shown in Fig. 9(a) for each superlattice, we can extract conventional XRD patterns. The measurement time of a concurrent XRD is about 100 times shorter than that of a conventional XRD. The results of simulating the diffraction patterns using one step model[30] is also shown in Fig. 9(b). The position of the diffraction peaks agrees well with the simulation results.

ACKNOWLEDGMENTS

The authors thank Prof. H. Koinuma, Prof. T. Hasegawa, Dr. T. Fukumura, Dr. T. Koida and Mr. D. Komiyama for useful discussions. This work was supported by the MEXT Grant of Creative Scientific Research (14GS0204), the MEXT Grant-in-Aid for Young Scientists (15685011), and a NEDO International Joint Research Grant (02MB3).

Figure caption

Fig. 1. (Color online) Illustration of the combinatorial laser molecular beam epitaxy system.

Fig. 2. (Color online) Schematics of combinatorial laser molecular beam epitaxy methods: (a) discrete cells, (b) parallel superlattice synthesis, (c) temperature gradient, and (d) composition-spread.

Fig. 3. (Color online) An (a) outside and (b) inside view of the combinatorial laser molecular beam epitaxy chamber..

Fig. 4. Schematic diagram of the control circuits for the combinatorial laser molecular beam epitaxy system.

Fig 5 (Color online)  (Top panel) RHEED intensity oscillations during the synthesis of $(SrTiO_3)/(BaTiO_3)$ superlattices. (Bottom panel) Schematic of the deposition sequence and the structure of the superlattices that were grown.

Fig. 6. (Color online) (a) Schematic side view of the scanning RHEED system when it is used in the rocking curve measurement mode. (b) Incident angle dependence of RHEED intensity oscillations during homoepitaxial growth of $SrTiO_3$.

Fig. 7. (a) Schematic diagram of the x-ray beam geometry in a concurrent x-ray diffractometer. (b) The Ewald construction of the concurrent XRD configuration in reciprocal space. (c) Photograph of the concurrent XRD system.

Fig. 8. (Color online) (a) Schematic illustration of a $Ba_xSr_{1-x}TiO_3$ composition-spread film grown on a $SrTiO_3$ (001) substrate. (b) A concurrent XRD image of the composition-spread film. The horizontal and vertical axes correspond to $x$ and the diffraction angle ($2\theta$). The colored contour lines denote logarithmic x-ray diffraction intensity. (c) The $x$ dependence of the out-of-plane and in-plane lattice constants of the composition-spread film.

Fig 9 (Color online) (a) Surface plot of diffraction intensity from a parallel library of ten $[(SrTiO_3)_m/(BaTiO_3)_m]_{30}$ (n = 12, 14, …, 28, 30) superlattices. The axes correspond to the position in the library and diffraction angle ($2\theta$). (b) Cross section of (a) for a m=14 superlattice together with a step model simulation.

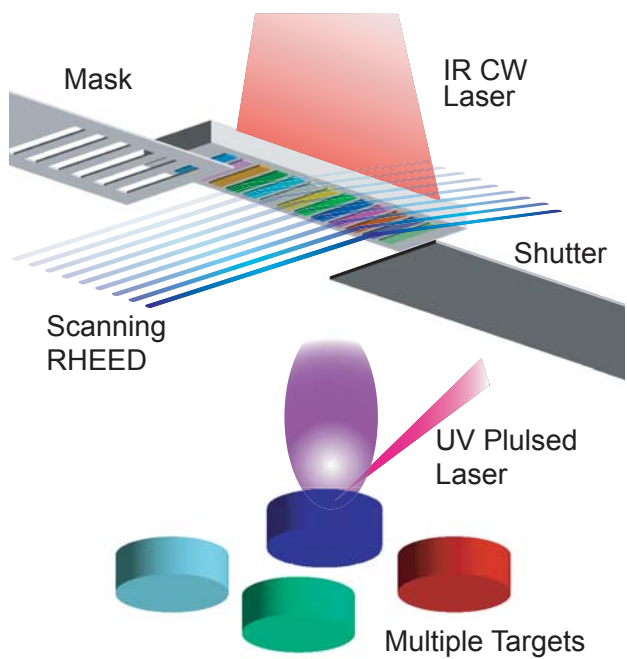

Fig. 1 M. Ohtani et. al.

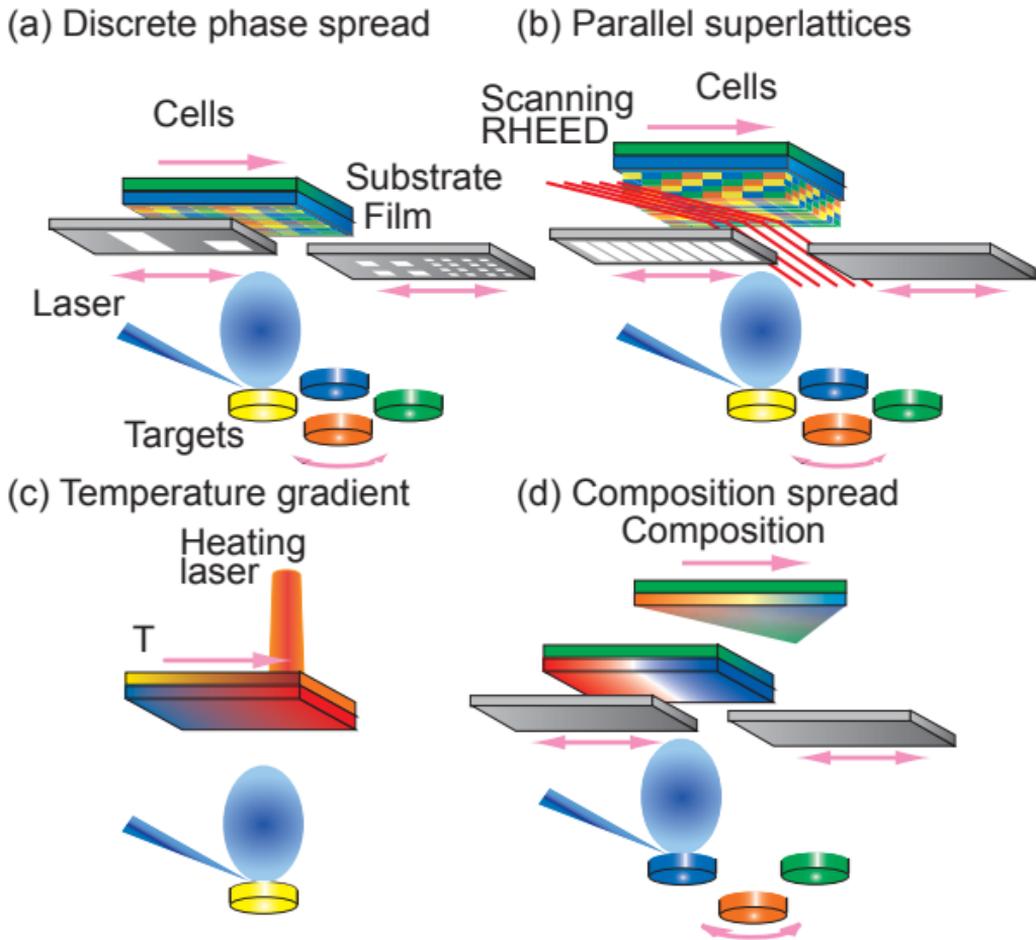

Fig. 2 M. Ohtani et. al.

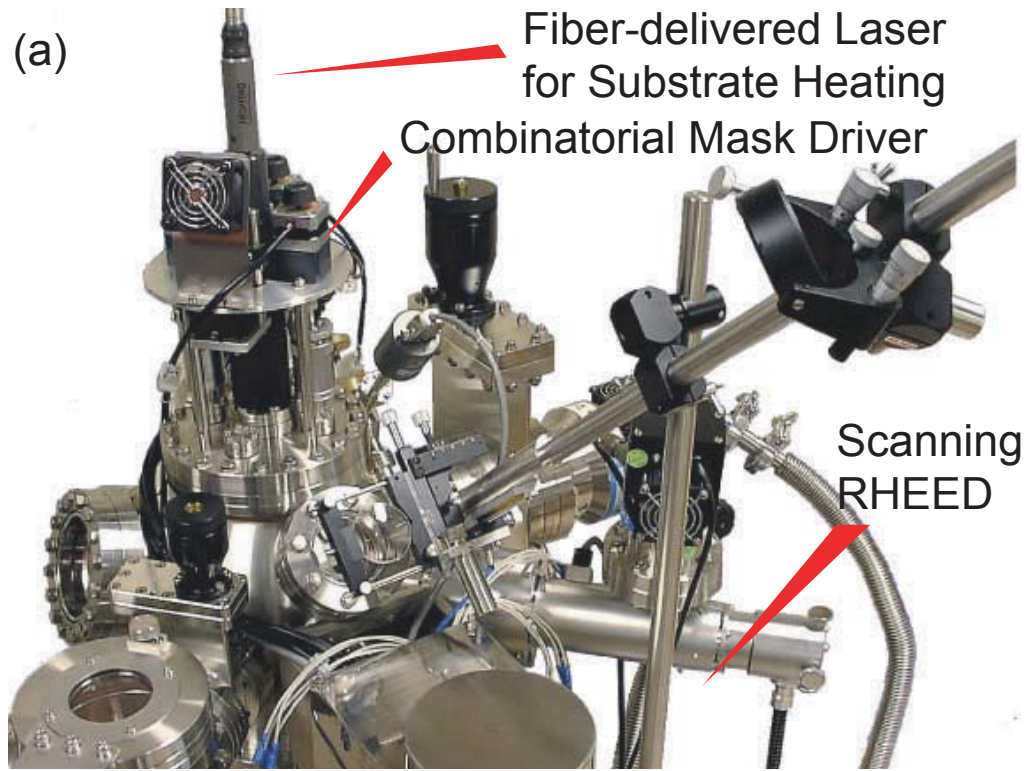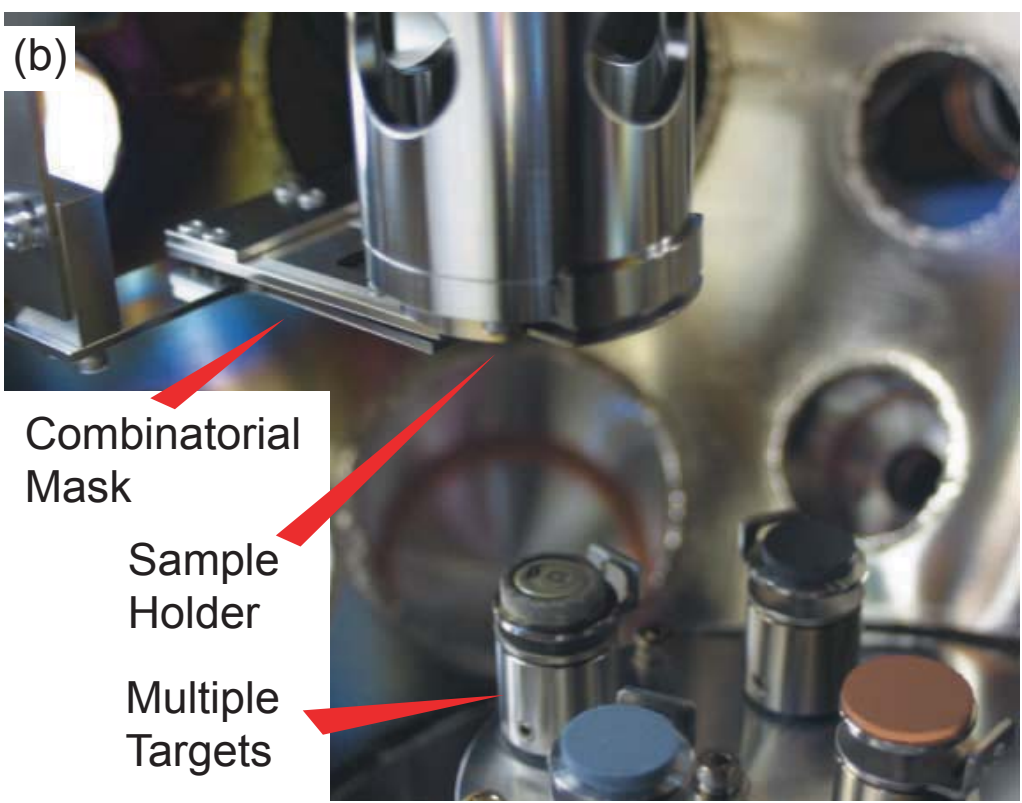

Fig. 3 M. Ohtani et. al.

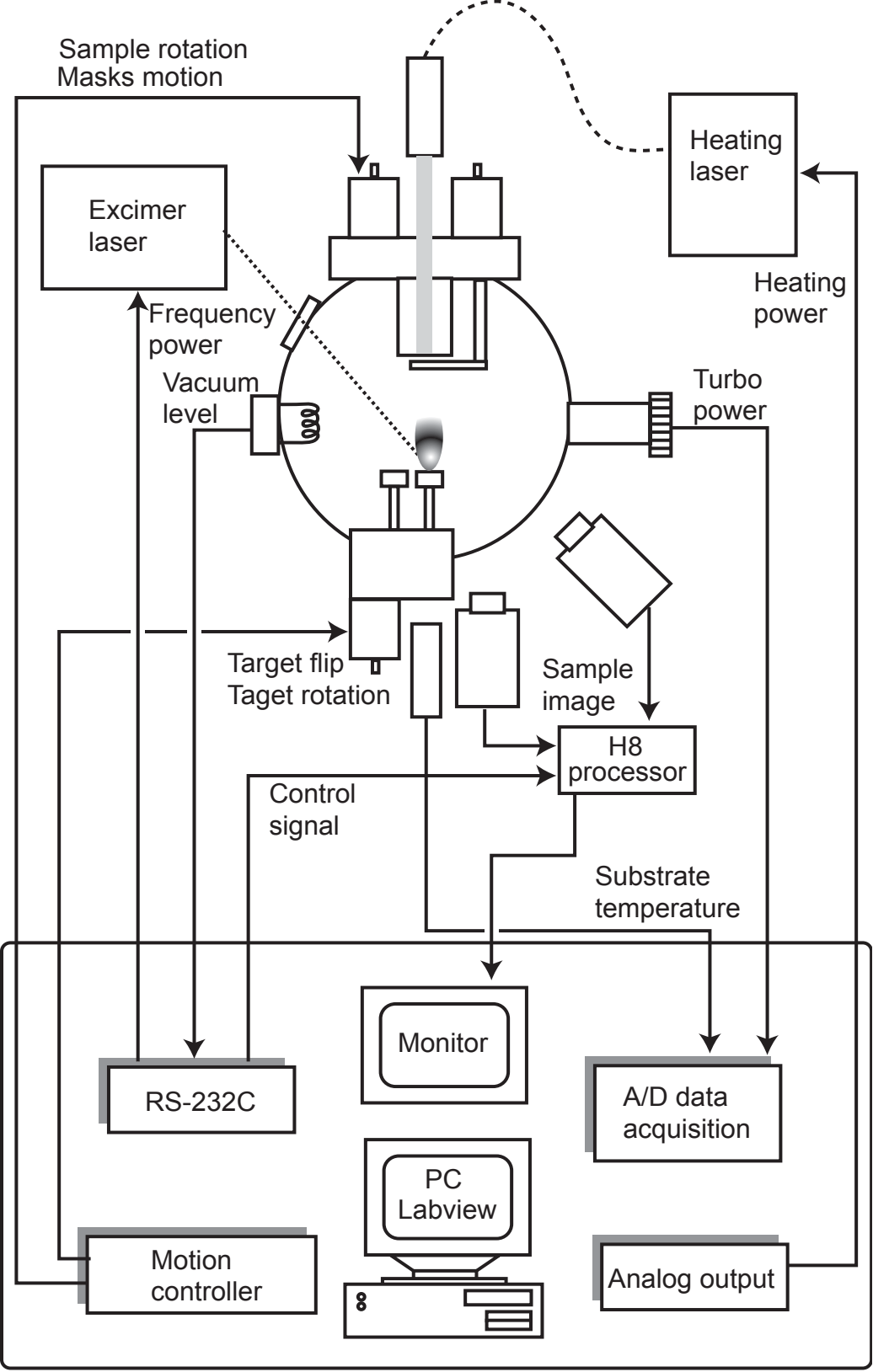

Fig. 4 M. Ohtani et. al.

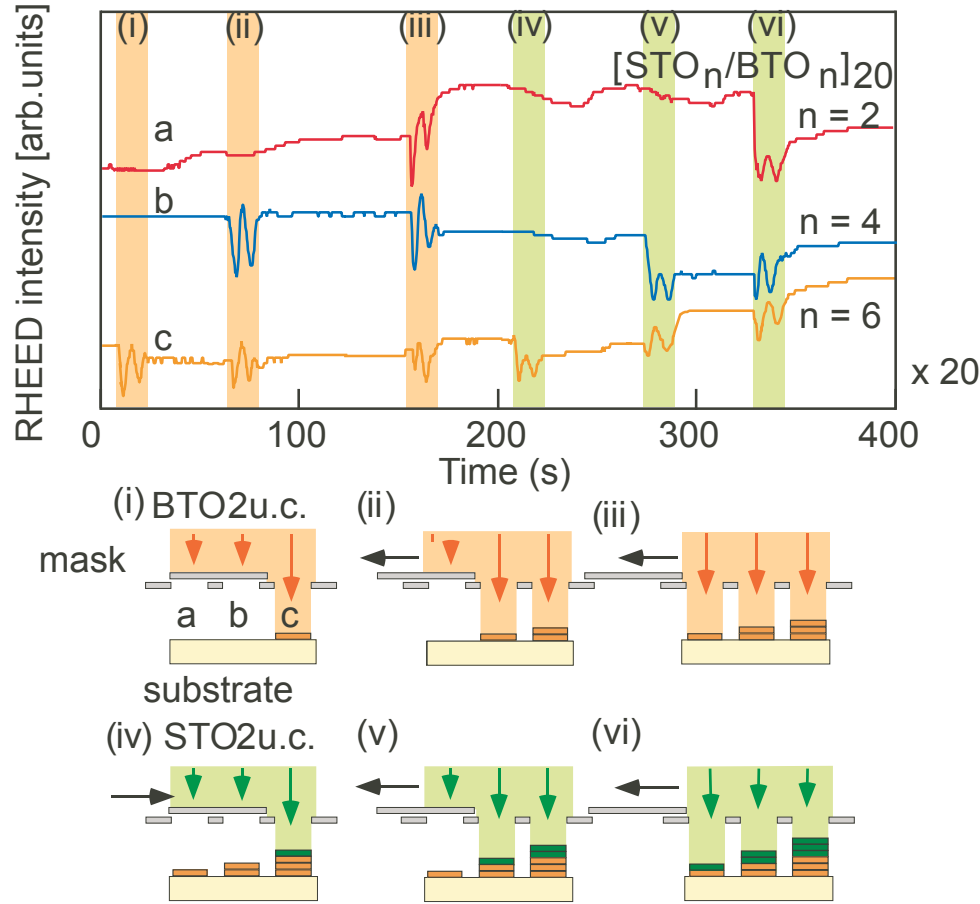

Fig. 5 M. Ohtani et. al.

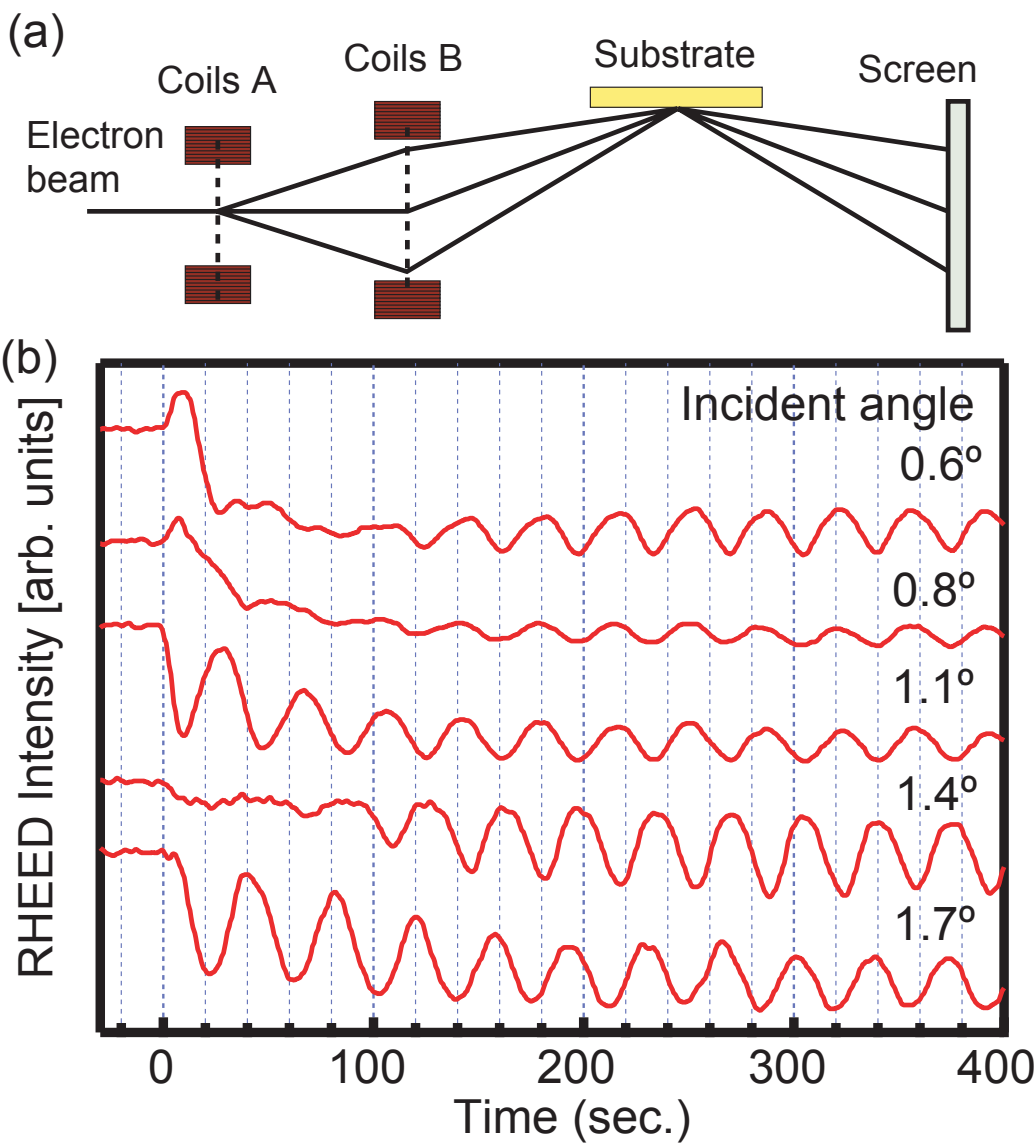

Fig. 6 M. Ohtani et. al.

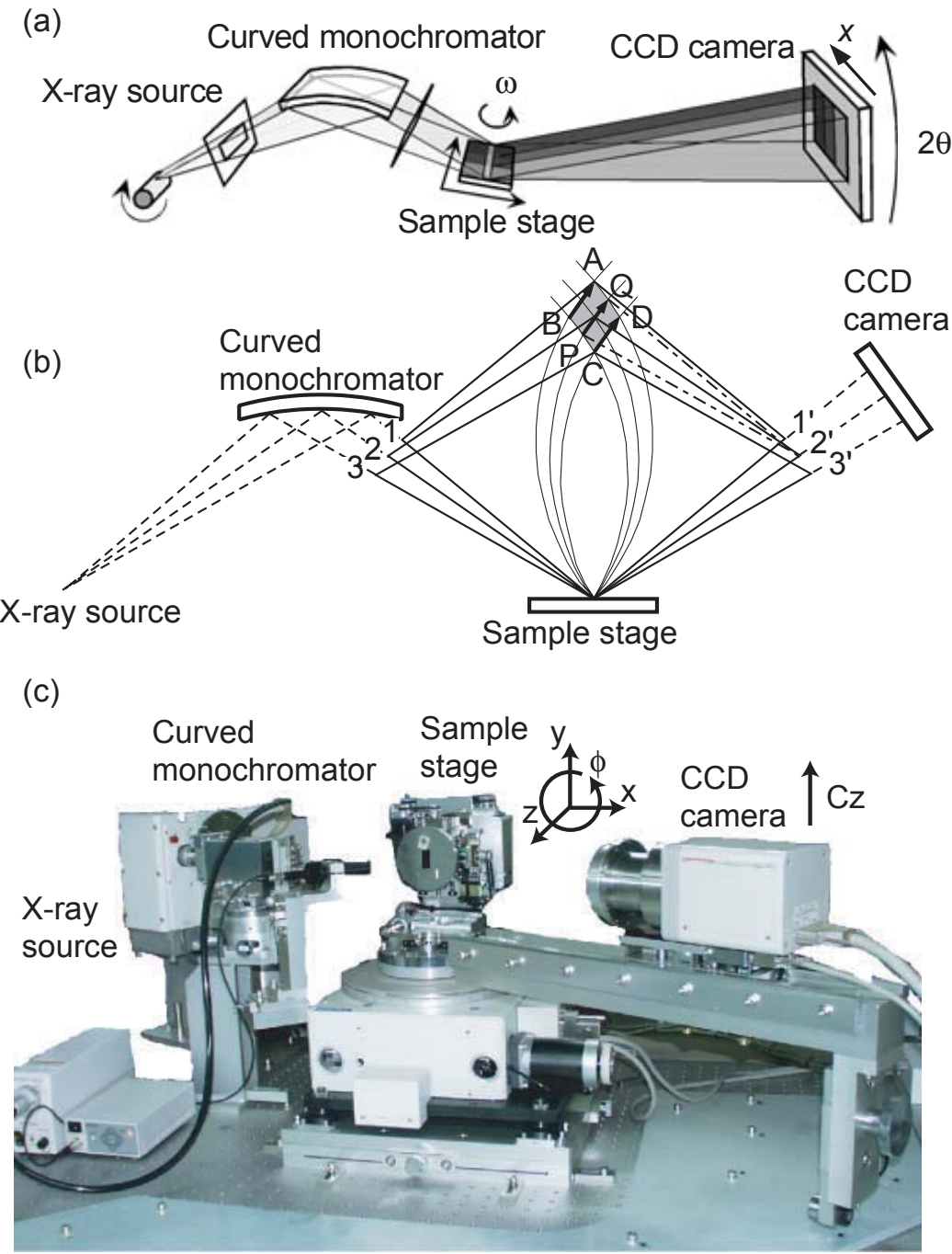

Fig. 7 M. Ohtani et. al.

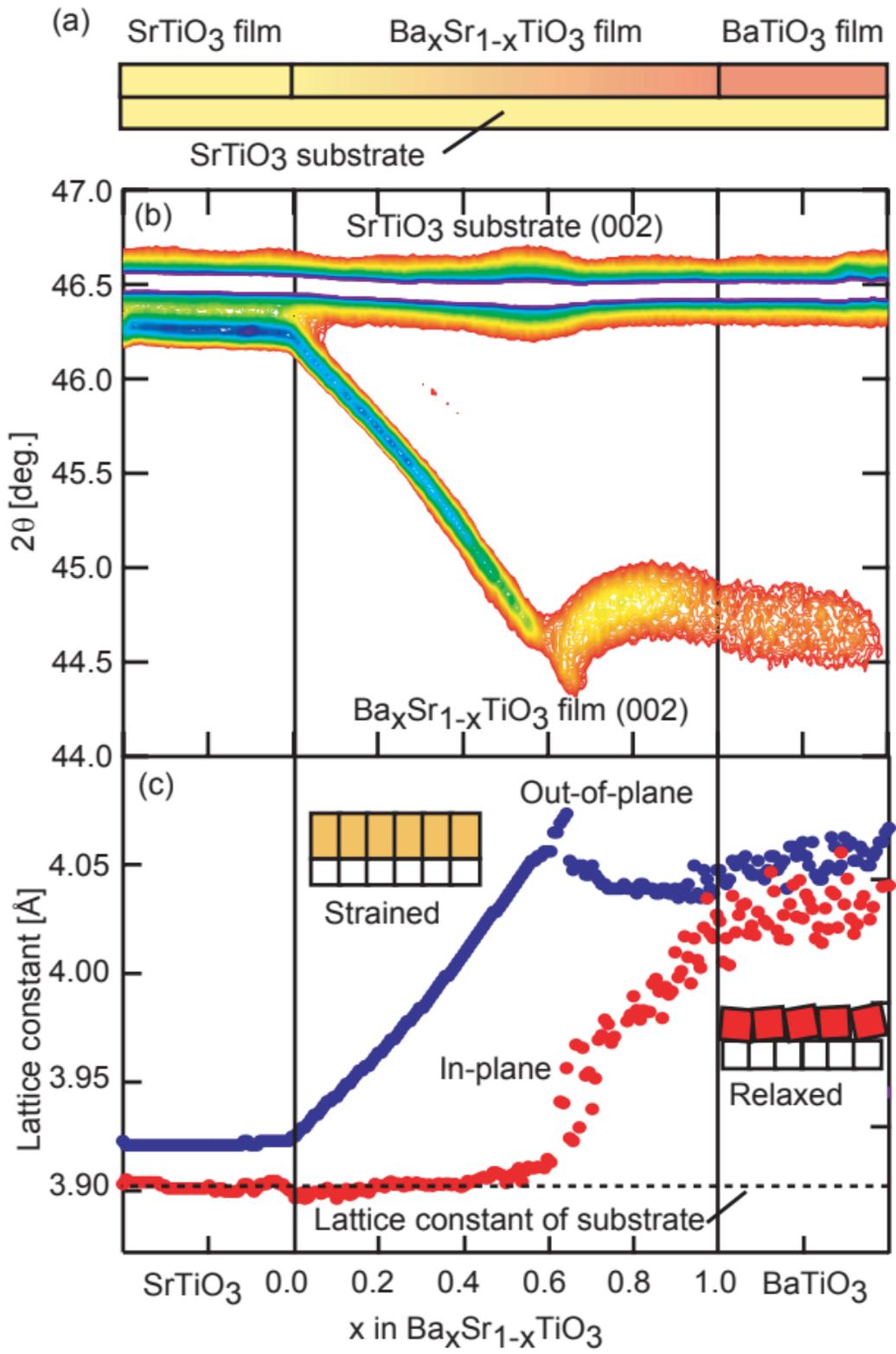

Fig. 8 M.Ohtani et. al.

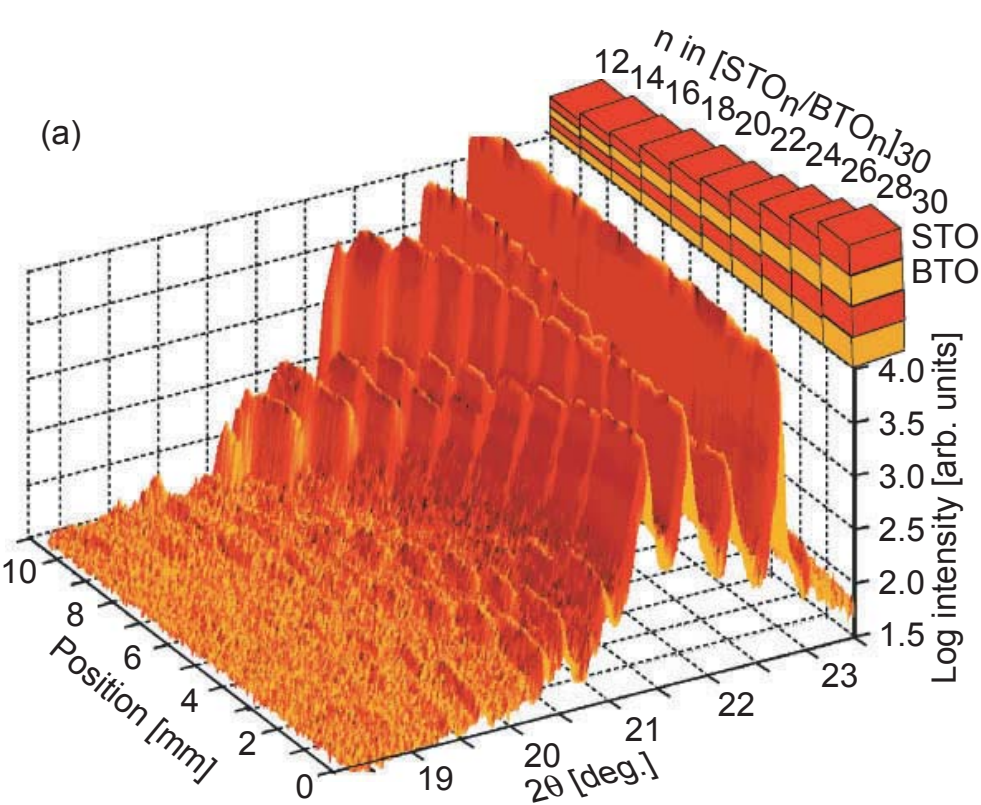

(a)

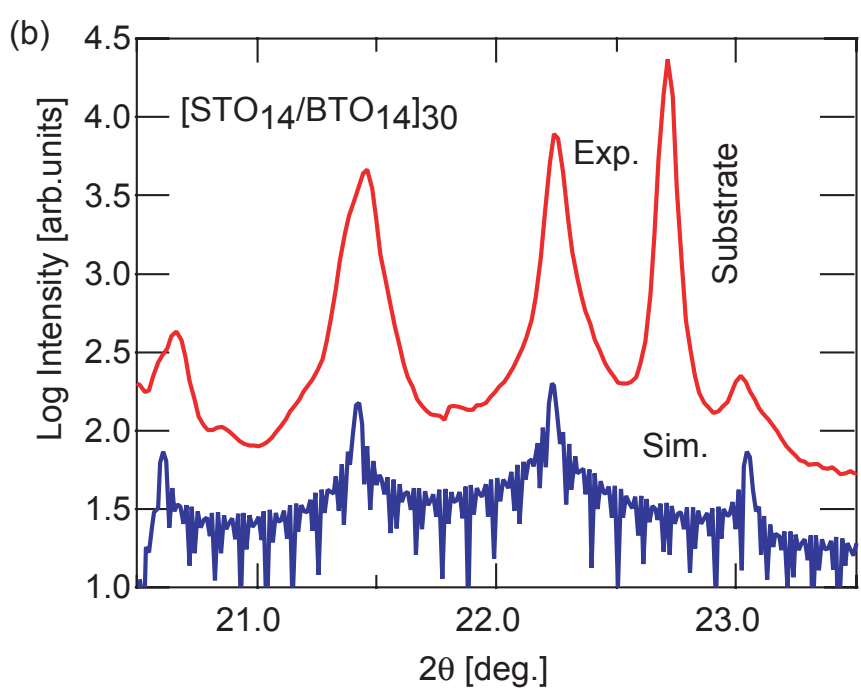

(b)

Fig. 9 M. Ohtani et. al.